\documentclass[apj]{emulateapj}

\shorttitle{BAO measurements on 21 cm surveys}
\shortauthors{Mao}

\begin{document}

\title{Measuring Baryon Acoustic Oscillations on 21 cm intensity fluctuations at moderate redshifts}

\author{Xiao-Chun Mao\altaffilmark{1} \altaffilmark{\dag}}

\altaffiltext{1}{National Astronomical Observatories, Chinese Academy of
Sciences, Beijing 100012, China}
\altaffiltext{\dag}{E-mail: xcmao@bao.ac.cn}

\slugcomment{The Astrophysical Journal, 752:80 (12pp), 2012 June 20}
\received{2011 December 28}
\accepted{2012 April 9}
\published{2012 May 29}

\begin{abstract}
After reionization, emission in the 21 cm hyperfine transition provides a direct probe of neutral hydrogen distributed in galaxies. Different from galaxy redshift surveys, observation of baryon acoustic oscillations in the cumulative 21 cm emission may offer an attractive method for constraining dark energy properties at moderate redshifts. Keys to this program are techniques to extract the faint cosmological signal from various contaminants, such as detector noise and continuum foregrounds. In this paper, we investigate the possible systematic and statistical errors in the acoustic scale estimates using ground-based radio interferometers. Based on the simulated 21 cm interferometric measurements, we analyze the performance of a Fourier-space, light-of-sight algorithm in subtracting foregrounds, and further study the observing strategy as a function of instrumental configurations. Measurement uncertainties are presented from a suite of simulations with a variety of parameters, in order to have an estimate of what behaviors will be accessible in the future generation of hydrogen surveys. We find that 10 separate interferometers, each of which contains $\sim 300$ dishes, observes an independent patch of the sky and produces an instantaneous field-of-view of $\sim 100$ $\rm deg^2$, can be used to make a significant detection of acoustic features over a period of a few years. Compared to optical surveys, the broad bandwidth, wide field-of-view and multi-beam observation are all unprecedented capabilities of low-frequency radio experiments.
\end{abstract}


\section{Introduction}

The length scale of baryon acoustic oscillation (BAO) provides a ``standard ruler'' in the Universe, making it one of the most powerful probes of cosmic expansion history. This characteristic scale is the distance traveled by sound waves in the photon-baryon plasma before recombination, and has been detected both in the cosmic microwave background (CMB) at $z \sim 1000$ \citep {Hinshaw09} and in galaxy surveys at $z\sim0$ \citep{Eisenstein05}. Recently, much effort has been expended to establish feasible schemes of measuring the BAO features accurately at a variety of redshifts. It is commonly believed that a combined analysis of BAO measurements and other independent observables can substantially improve the current constraints on the equation of state of dark energy and its time dependence \citep[e.g.][]{Seo03,Hu03,Zhan08,Zhang09,Samushia09,Wang11}.

Since emission in the 21 cm hyperfine transition can be safely assumed to be proportional to the cosmic density of neutral hydrogen, measurement of intensity fluctuations in the cumulative 21 cm emission appears as a very promising technique to trace large scale structure (LSS) over a wide range of redshifts. In this manner, BAO signatures may be detected with small errors and used to gain cosmological insight \citep[e.g.][]{Wyithe08,Mao08,Chang08,Ansari08,Rhook09}. Because the structure of interest is certainly above the non-linear scale and the 21 cm radiation is sensitive to neutral hydrogen regardless of whether it is part of a resolved object, fine angular resolution is not necessary for such experiments. Thus the blind 21 cm surveys could be more efficient than traditional galaxy surveys at high redshifts where individual sources are hard to be resolved in sufficient numbers. Moreover, the 21 cm experiments have the advantage of simultaneously exploring a large redshift range, benefiting from the broad frequency bandwidth allowed by radio telescopes. Furthermore, the 21 cm emission line can provide a direct and precise measure of the redshift, which is unlikely to be true in galaxy surveys. A considerable survey volume is necessary for 21 cm measurements to achieve the precision required to detect large-scale features. Cylindrical radio telescopes with a wide field of view have been proposed to perform such surveys in the future \citep{Peterson06,Seo10,Ansari11}, and near term measurements using existing telescopes are discussed by \citet{Masui10}.

%Fig.1
\begin{figure*}
\begin{center}
\includegraphics[angle=270, scale=0.6]{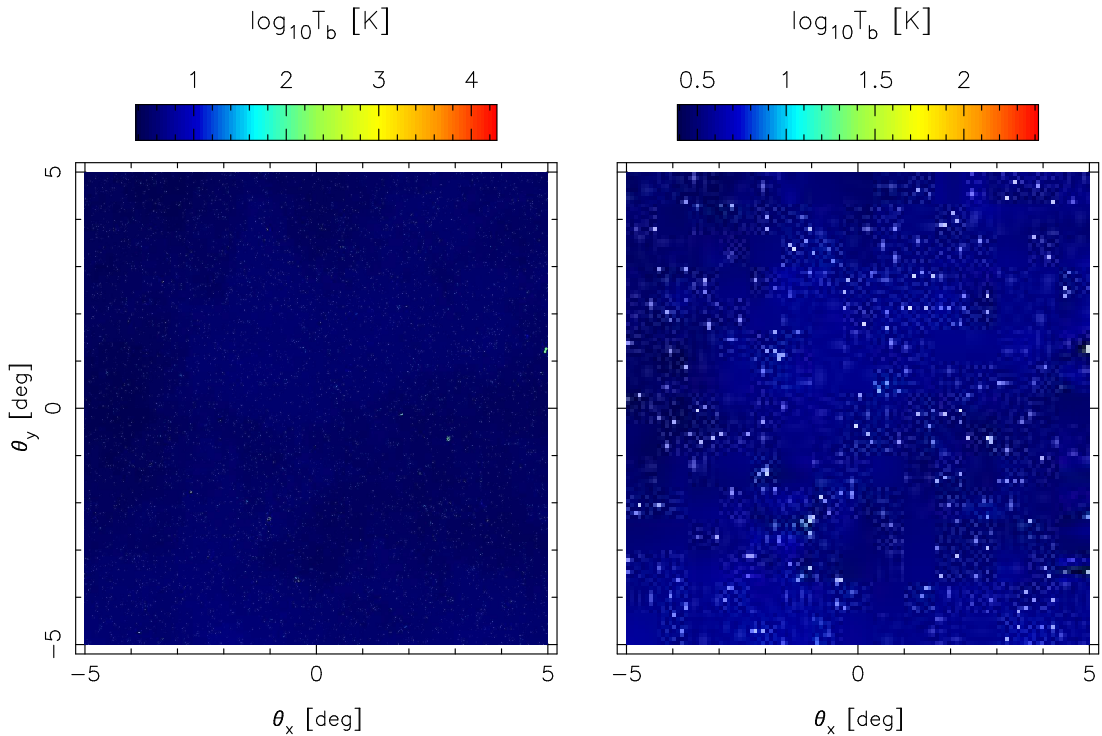}
\caption{Brightness temperature images of foregrounds simulated at $\nu=710$ MHz. The angular resolutions are nearly 0.6 arcminutes (left panel) and 5 arcminutes (right panel).}
\end{center}
\end{figure*} 

However, all of the most important LSS surveys have been taken with optical but not radio telescopes. BAOs are weak features imprinted in the 21 cm cosmological signal which has a strength of $\sim10$ $\mu$K near $z\sim1$. In addition to the extremely faint signals themselves, foreground contamination seems formidable in low-frequency experiments. The Galactic and extragalactic foreground sources exceed the cosmological signal by at least four orders of magnitude \citep{Bowman09,Liu09a,Chang10,Wang10}. Radio techniques have so far yielded direct detections of the 21 cm emission only to $z=0.24$ \citep{Lah07}, indicating a significant uncertainty in future neutral hydrogen surveys. The ability to constrain the nature of dark energy represents the ability to detect BAO wiggles in the 21 cm power spectrum. Keys to these detections lie not only in experimental strategies but also in data processing techniques. In this paper, we discuss the detectability of BAOs in future hydrogen surveys around $z=1$. Simulations of the 21 cm interferometric measurements are described, and a probable method of foreground subtraction is also considered. We present predictions for possible systematic and statistical errors in the BAO scale estimates over an important parameter range, showing the range of behaviors we might expect in future observations. We note that the foreground strategy in the post-reionization era is very different from that in the epoch of reionization (EoR). At high redshifts ($z>6$), most of the neutral hydrogen is distributed in intergalactic space, nearly separated from galaxies. Hence the extragalactic point sources can be treated as just foreground contaminants. However, the neutral hydrogen is locked in the overdense regions such as gas-rich galaxies at medium redshifts, indicating that the redshifted 21 cm line will be emitted from the extragalactic point sources, like the continuum. Galaxies therefore play two completely different roles here: foreground sources as well as cosmological signal emitters. When we consider the foreground subtraction technique at $z\sim1$, it must be kept in mind that only the continuum of galaxies should be cleaned out. Efficient operations should not remove foregrounds at the sacrifice of desired signals. We can neither simply discard the contaminated pixels in sky maps as suggested in CMB observations \citep{Tegmark98} nor remove point sources with the prior sky models as discussed in reionization experiments \citep{Datta10}. Recently, analyses of radio observations at higher frequencies (e.g., 610 MHz) have provided some possible ways to separate the post-reionization 21 cm signal from astrophysical foregrounds \citep[e.g.][]{Chang10,Ghosh11a,Ghosh11b}. And our work can be used as a starting point for modeling BAO detections in blind 21 cm surveys up to $z\sim5$.

This paper is organized as follows. In Section 2, we introduce the proper simulations of the foreground model as well as the radio interferometric measurements. Our foreground removal technique is also described. Next, we apply the subtraction method to the simulated data cube in Section 3, and further estimate the sensitivities on measurements of the 21 cm power spectrum. We analyze results from a series of simulations in order to understand how the measurement uncertainties depend on the data reduction and instrumental configurations. Finally, we summarize our main conclusions in Section 4. Throughout the paper we adopt a concordance cosmology of $\Omega_0=0.265$, $\Omega_\Lambda=0.735$, $\Omega_b=0.044$, $h=0.71$, $n_s=1$ and $\sigma_8=0.772$, as revealed by the WMAP three-year observations \citep{Spergel07}.

%Fig.2
\begin{figure*}
\begin{center}
\includegraphics[angle=270, scale=0.6]{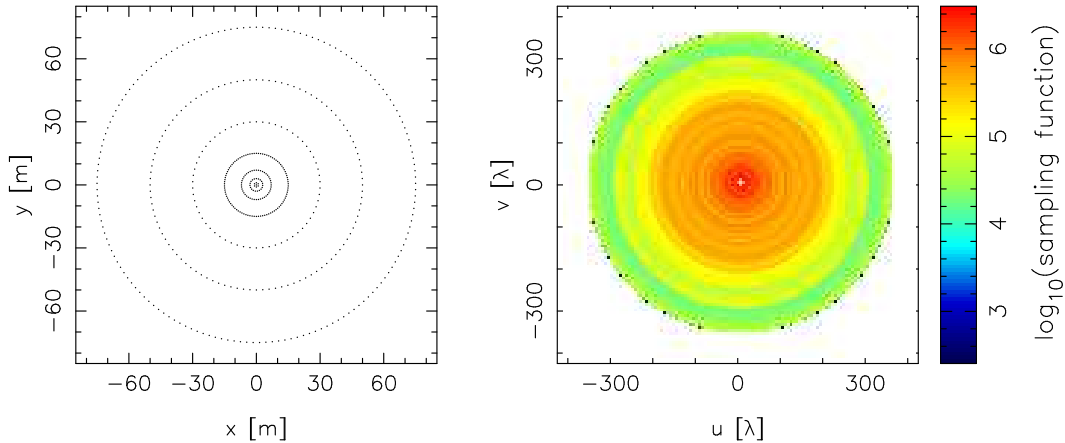}
\caption{Arrangement of antenna elements in the radio interferometer (left) and the corresponding density of visibility measurements in the \emph{uv} plane at 710 \rm{MHz} with integration time of 24 hours (right). All the 487 dishes are involved, distributed in 7 circles with diameters of $D=2,6,14,30,60,100,150$ m.}
\end{center}
\end{figure*}

\section{Method}

\subsection{simulation of foregrounds}

We employ Monte-Carlo simulations presented by \citet{Wang10} and references therein
to generate the brightness temperature sky maps at relevant frequencies, which incorporate contributions from three main components: (1)Galactic synchrotron and free-free emission; (2)galaxy clusters; and (3)extragalactic discrete sources such as star-forming galaxies and AGNs. Random variations of morphological and spectroscopic parameters as well as the evolution of radio halos in galaxy clusters are also considered. The outline of these foreground simulations has been given in \citet{Mao11}, and more details can be found in the original papers.

All the simulations are performed on $1024\times1024$ grids over a field-of-view of $10^{\circ}\times10^{\circ}$. Our foreground box is kindly provided by \citet{Wang10}. Since most BAO experiments are designed to provide low angular resolutions, we further arrange the sky maps onto grids of $120\times120$ pixels, corresponding to the angular resolution of 5 arcminutes. Along the line of sight, we use the frequency range extending from 500 MHz to 900 MHz with a resolution of 1 MHz. In Figure 1, the simulated foreground maps with high and low angular resolutions are shown separately.

%Fig.3
\begin{figure*}
\begin{center}
\includegraphics[angle=270, scale=0.7]{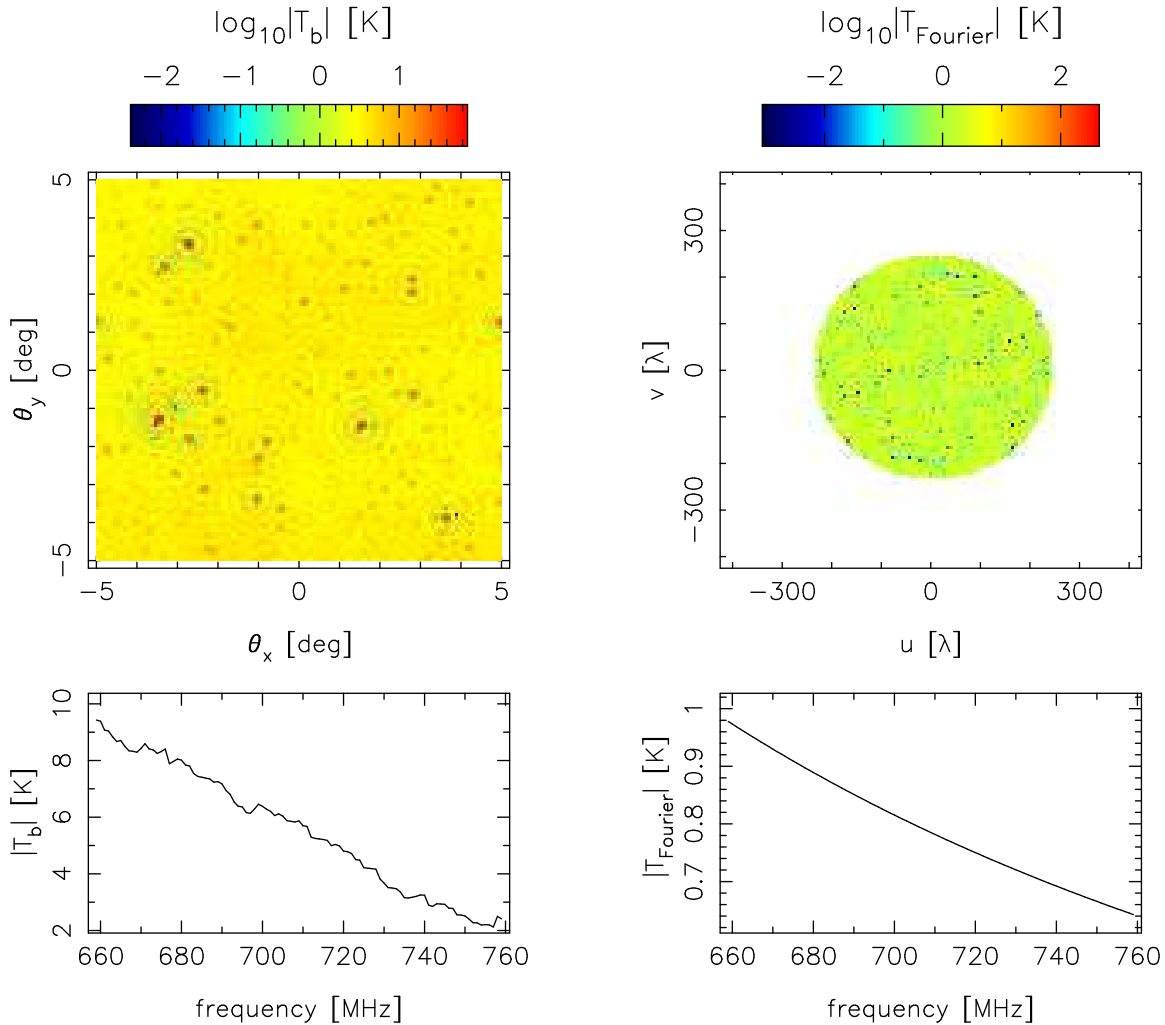}
\caption{The top row shows the dirty map (left panel) and the corresponding \emph{uv} map (right panel) observed at $\nu=710$ MHz under the fiducial model, while the bottom row shows the frequency spectra in a random image pixel (left panel) and \emph{uv} pixel (right panel) across the frequency range $660\le\nu\le760$ \rm{MHz}.}
\end{center}
\end{figure*}

\subsection{simulation of radio interferometric measurements}

To calculate the sensitivity of a BAO measurement, we need to specify both the instrumental configurations and the observing strategy. Many factors contribute to the instrumental response of a radio interferometer. Among the most important to consider are the field of view, angular resolution, collecting area, antenna distribution, and bandwidth \citep{Bowman06}.

In this work, the smallest scale we consider is set to be the third BAO peak which locates at $\sim 0.5$ degree angular scale around $z=1$ \citep{Ansari11}. And thus a radio instrument of modest size, typically with diameter $D\sim 100$ m, can be used to resolve the BAO features. Moreover, BAO surveys should be performed in various pointing directions in order to observe a large volume of the universe which is crucial to obtain small statistical errors on measurements of the LSS power spectrum. Hence the characteristic observing time scales with the collecting area, the instantaneous field of view (\emph{i.e.}, FOV) and the number of pointings. Obviously, a large FOV $\sim(\lambda/D_{\rm dish})^2$ is important to raise the survey efficiency. Here, we introduce an interferometer consisting of $N_{\rm dish}=487$ dishes with diameter $D_{\rm dish}=1$ m. All the dishes are nearly coplanar and organized in circles with different diameters of $D=2,6,14,30,60,100,150$ m. For $D\leq30$ m, the dishes are spaced roughly 1 m from each other, otherwise, the dishes are spaced roughly 3 m from each other. In the left panel of Figure 2, we show the array layout for the 487 dishes with a maximum baseline of 150 m. The corresponding \emph{uv} sampling density at $\nu=710$ MHz is plotted in the right panel. One may note that, when the dishes are limited in quantity, the dense-to-sparse arrangement could promise complete \emph{uv} coverage in the central area, but also provide some long baselines for good angular resolutions.

%Table 1
    \begin{deluxetable}{lr}
    \tablewidth{18pc}
    \tablecaption{Experimental Specifications}
%    \caption{Experimental Specifications}

%        \label{tab_detector}
    %}

    \tablehead{\colhead{Parameter} & \colhead{Value}}

    \startdata
    $\Omega_{\rm HI}b$                              & 0.0005\\
    Array diameter, D (m)                           & 100\\
    Number of dishes, $N_{\rm dish}$                & 330 \\
    System temperature, $T_{\rm sys}$ (K)           & 50 \\
    Bandwidth, B (MHz)                              & 100 \\
    Central frequency, $\nu_0$ (MHz)                & 710 \\
    Spectral resolution, $\Delta \nu$ (MHz)         & 1 \\
    Order of polynomial fit, N                      & 3 \\
    Sky coverage, $f_{\rm sky}$                     & 0.5 \\
    \enddata

    \tablecomments{ Parameters are listed for the fiducial model. The central frequency corresponds to $z=1$ in 21 cm measurements.}
    \end{deluxetable}

How shall we carry out the large-volume survey? Generally speaking, a larger sky coverage requires a longer observation time. However, if we could observe several independent fields of view simultaneously, the total survey volume will be extended without prolonging the observing time. Here we consider a 21 cm survey with 10 separate interferometers, each situated at a different location and observing a different patch of the sky. Compared to a single interferometer, the cost of building the proposed telescope will no doubt increase. And we need to find a balance between price and efficiency. For computational simplicity, our reference measurement is chosen to be a deep observation of a single target field centered on the North Celestial Pole (NCP). The effect of the earth rotation synthesis is not taken into account in the reference observation. We assume that observations pointing in different directions can be performed in the same way as our reference measurement.

%Fig.4
\begin{figure*}
\begin{center}
\includegraphics[angle=270, scale=0.6]{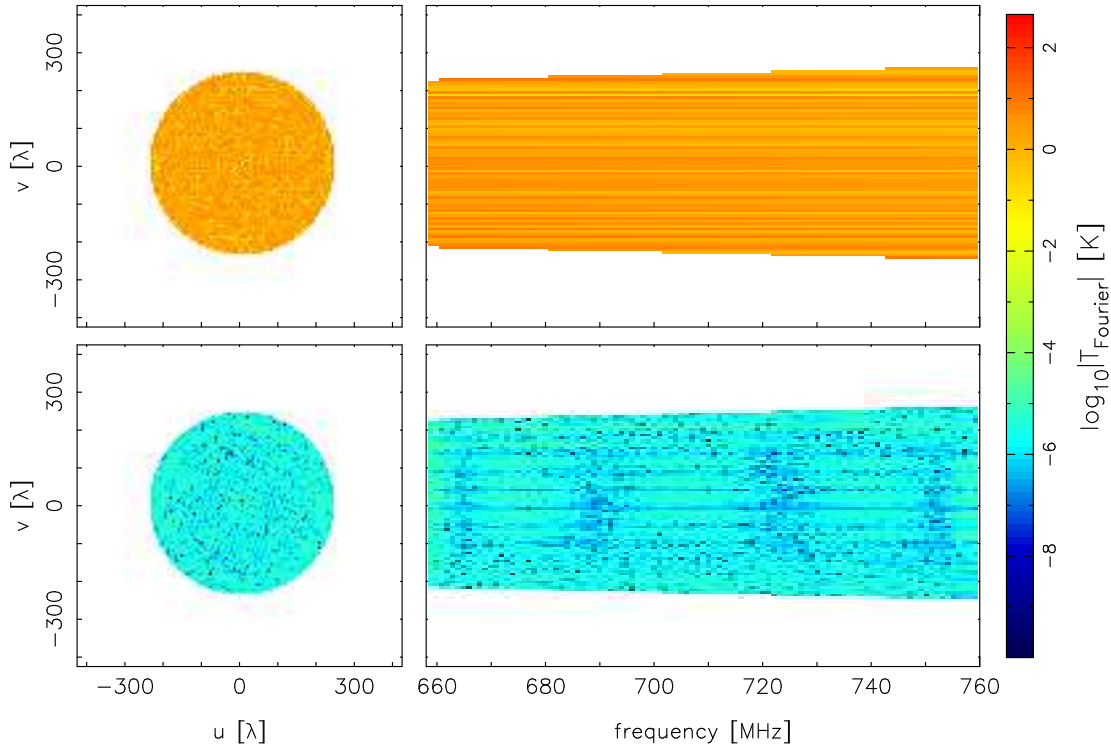}
\caption{Foreground subtraction technique applied to the simulated interferometric measurements. The top row gives two cuts for the input visibility cube, showing a \emph{uv} map at a single frequency channel $\nu=710$ \rm{MHz} (left panel) and a slice along the frequency direction and \emph{v}-axis (right panel). The bottom row gives the same cuts, but for the cleaned visibility cube following foreground subtraction.}
\end{center}
\end{figure*}

%Fig.5
\begin{figure}
\begin{center}
\includegraphics[angle=270, scale=0.5]{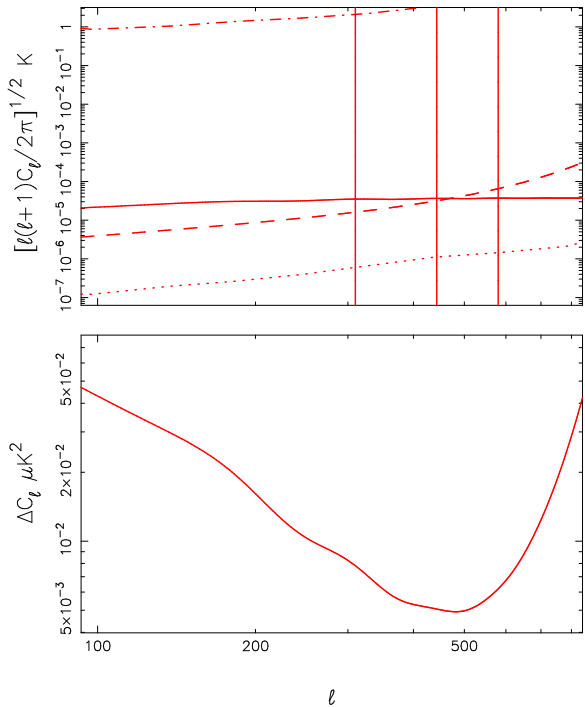}
\caption{Angular power spectra and measurement variance at $z=1$ for the fiducial model. Top panel: the dash-dotted line and dotted line correspond to the input foregrounds and detector noise, respectively. The solid line represents the expected 21 cm signal, and the dashed line shows the post-subtraction noise in the cleaned sky map, including residual foregrounds and detector noise. The vertical lines indicate locations of the first three BAO peaks. Bottom panel: variance in the estimate of 21 cm power spectrum for an integration time of 1 year.}
\end{center}
\end{figure}

In order to understand how the instrumental parameters impact the future BAO measurements, our simulations are passed through the observational pipeline described in \citet{Mao11}. In what follows, we briefly introduce its scheme. Since the 21 cm signal is very faint compared to foregrounds and its low-frequency properties are highly uncertain, we do not include it in our sky model. Besides, the effect of foreground removal on the cosmological signal is outside the scope of this work and hence ignored. The original image cube is first established on the basis of the brightness temperature distribution of foreground sources. Also, the primary beam is simply assumed to be unity within the FOV. At each frequency channel, the sky image is related to visibilities via the two dimensional Fourier transform. Subsequently, we estimate the noise visibilities with one year integration. The rms noise per visibility per frequency channel can be written as \citep{Rohlfs04,McQuinn06}
\begin{eqnarray}
\Delta V_{\nu}^{N}(u,v)=\frac{\lambda^2 T_{\rm sys}}{A_e\Omega_b\sqrt{\Delta \nu t}},
\end{eqnarray}
in which $A_e$ and $\Omega_b$ are the effective area and the beam solid angle of a single dish respectively, $\Delta \nu$ is the bandwidth of a single frequency channel, and $t$ is the total integration time for sampling a given $(u,v)$ location. For the reference observation, we assume the system temperature to be $T_{\rm sys} = 50$ \rm{K} according to current technology. In addition, we approximate the integration time $t=\tau N(u,v)$, where $\tau=5$ \rm{s} is the accumulation duration for each visibility measurement, and $N(u,v)$ is the number of independent samples in that pixel \citep{Bharadwaj05,Bowman09}. Because the thermal noise is random, we draw complex visibilities from Gaussian distributions with zero mean and rms described above. Finally, the real-world sampling is carefully performed in the \emph{uv} plane. The contribution of any single visibility measurement is applied to only one grid cell. We further normalize the baseline distribution to ensure that each pixel in the sampled part of the \emph{uv} plane has the same weight. With the uniform weighting, we artificially emphasize the information contained in long baselines and increase the effective resolution of the derived sky maps.

Following all these simulations, we generate our image cube and visibility cube representing realistic observations. Figure 3 shows the dirty map and the corresponding \emph{uv} map at $\nu=710$ MHz. Additionally, the frequency spectra in a random image and \emph{uv} pixel are also plotted separately in the bottom row. As illustrated in these two panels, the frequency coherence of foregrounds will be destroyed in real space owing to the ``mode-mixing'' effect \citep{Bowman09,Liu09a}.

%Fig.6
\begin{figure*}
\begin{center}
\includegraphics[angle=270, scale=0.7]{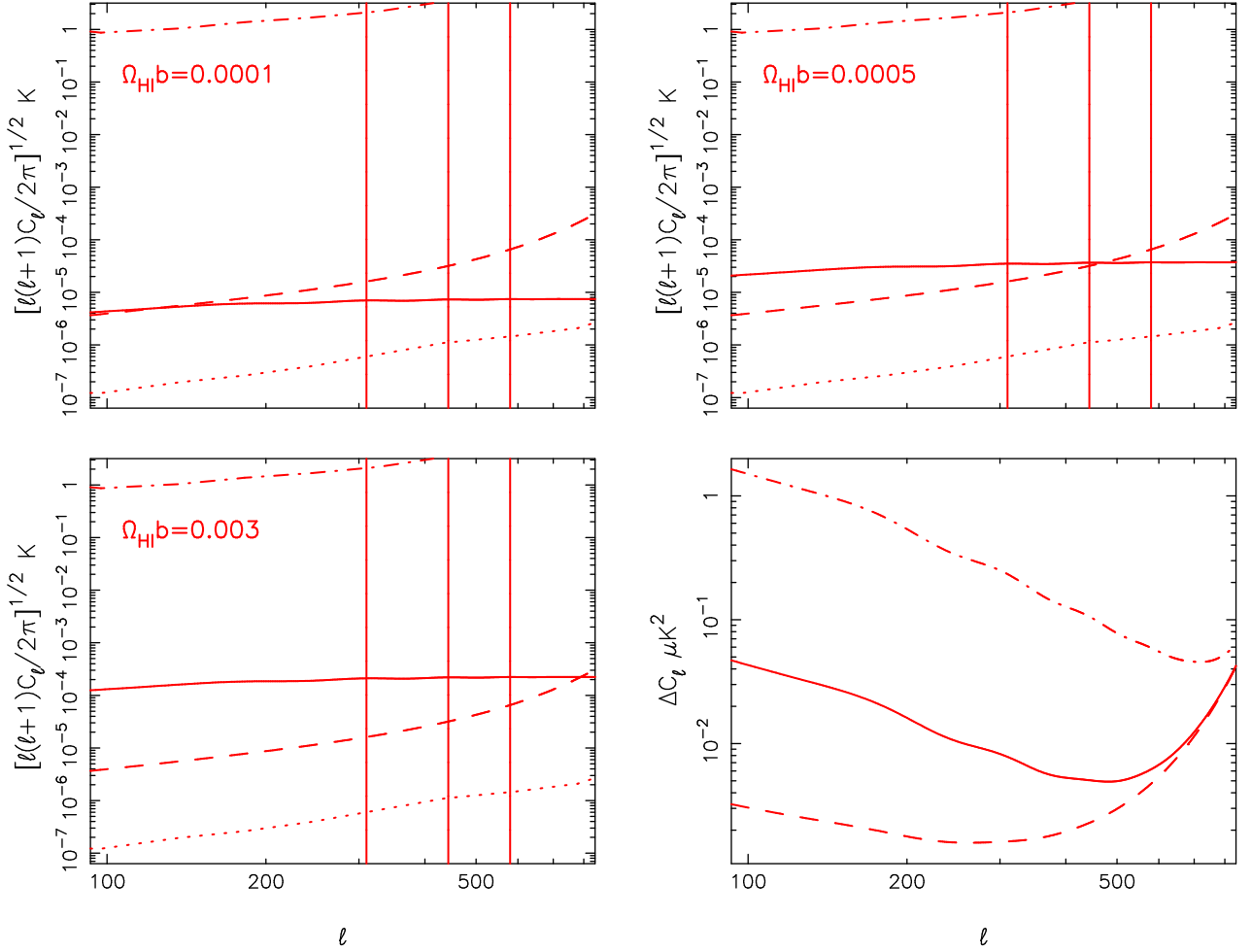}
\caption{Effect of the cosmic density of neutral hydrogen together with the bias factor: sensitivity estimates for the fiducial model but with various values of $\Omega_{\rm HI}b$. The top row as well as the bottom left panel give the angular power spectra as defined in the top panel of Figure 5. The bottom right panel shows the possible variances for $\Omega_{\rm HI}b=0.0001$ (dashed line), $\Omega_{\rm HI}b=0.0005$ (solid line), and $\Omega_{\rm HI}b=0.003$ (dash-dotted line).}
\end{center}
\end{figure*}

\subsection{foreground subtraction}

It is generally agreed that the intergalactic medium (IGM) contains the majority of baryons in the Universe, and is completely ionized by UV radiation beyond $z=6$ \citep{Becker01,Fan06}. Nevertheless, the damped Lyman series absorption seen in quasar optical spectra indicates that a significant fraction by mass of neutral hydrogen does reside in the interstellar medium (ISM) of galaxies \citep{Storrie00,Peroux01}, making it possible to use the neutral hydrogen as a tracer of the underlying large-scale structure in the post-reionization epoch. As we mentioned above, most of the neutral hydrogen is outside and inside of galaxies before and after reionization respectively, indicating that new developments in methods of isolating the 21 cm signal from foregrounds will be required in post-reionization experiments. In such observations, emissions of galaxies consist of two components: the radio continuum as well as the redshifted 21 cm line. The former is relatively featureless, however, the latter is expected to fluctuate rapidly in the frequency and angular directions. Both these radiations are emitted from extragalactic point sources, but have fundamentally different qualities. In addition, the instrumental response changes with observing frequency, and thereby creates different sidelobe patterns across the measured sky maps, inducing the ``mode-mixing'' effect which will destroy the foreground smoothness in frequency space. The frequency decoherence has been shown clearly in the left hand column of Figure 3.

%Fig.7
\begin{figure*}
\begin{center}
\includegraphics[angle=270, scale=0.7]{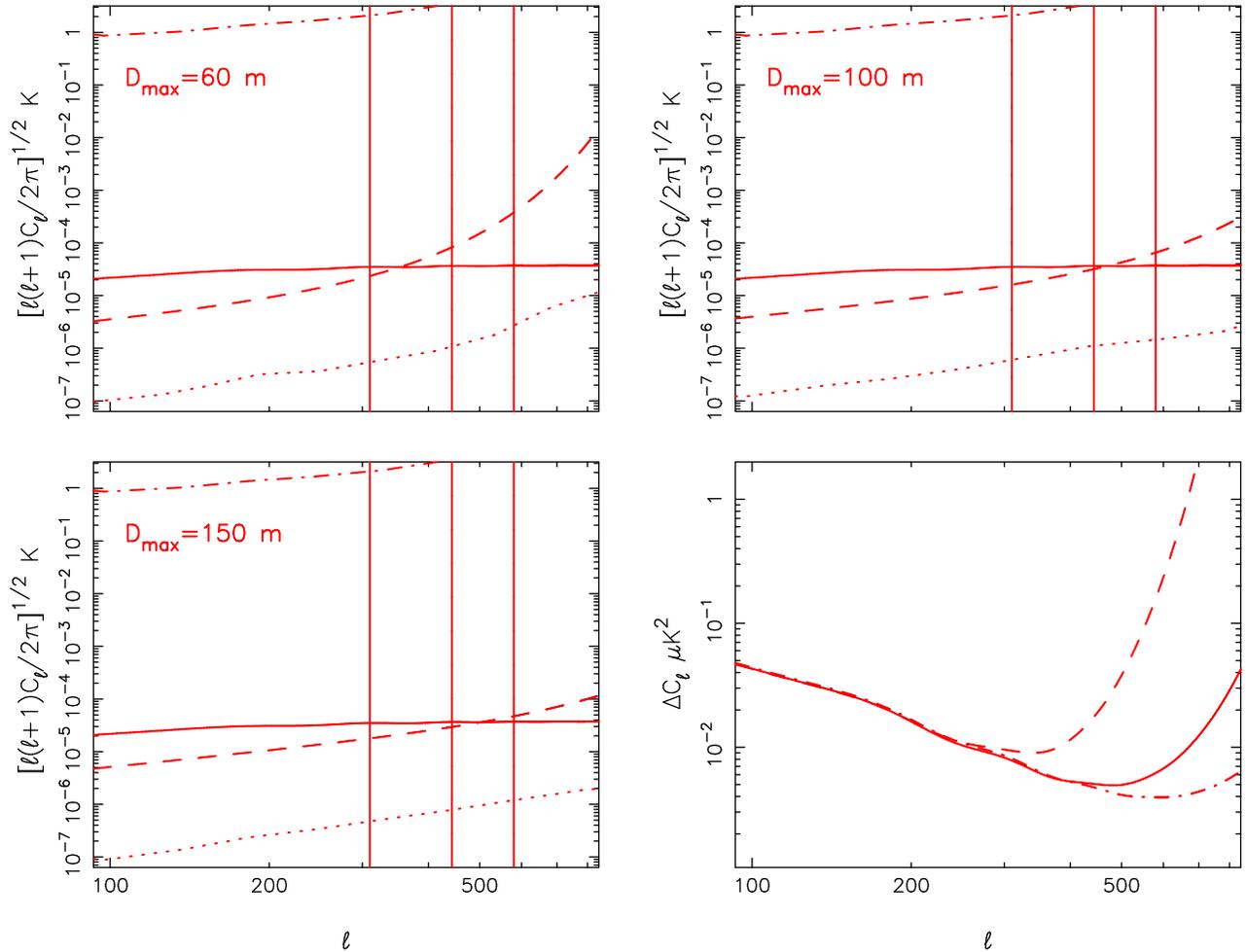}
\caption{Same as Figure 6, but for the effect of the array layout: sensitivity estimates for the fiducial model but with various arrangements of dishes. From the top left, top right to bottom left panel, the maximum baselines are $D_{\rm max}=60,100,150$ m respectively, and the corresponding total numbers of dishes are $N_{\rm dish}=226,330,487$. The bottom right panel shows the possible variances for $D_{\rm max}=60$ m (dashed line), $D_{\rm max}=100$ m (solid line), and $D_{\rm max}=150$ m (dash-dotted line).}
\end{center}
\end{figure*}

Over the last decade, much effort has been made in exploring possible methodologies for foreground subtraction \citep[e.g.][]{Matteo02,Oh03,Zaldarriaga04,Furlanetto04,Santos05,Wang06,Morales06,Gleser08,
Bowman09,Liu09a,Liu09b,Harker09,Harker10,Liu11,Petrovic11,Ghosh11a,Ghosh11b,Ansari11,Chapman12}. The most widely discussed technique concentrates on the spectral fitting along each line-of-sight (LOS), dealing with the removal of diffuse emission and unresolved point sources under the assumption that bright point sources have been cleaned out with high accuracy and precision. Owing to the ``mode-mixing'' effect, excision of point sources is always first carried out in a cleaning process \citep{Morales06,Datta09,Datta10,Pindor11,Bernardi11}. However, for the proposed neutral hydrogen surveys, the extragalactic point sources should be treated not only as foreground contaminants but also as signal emitters. This means that the point sources should not be removed, except for their continuum. \citet{Mao11} introduced a foreground removal technique in Fourier space, which handles the confusion-level contaminants and the bright point sources as equivalent. They fit the visibility spectrum using a weighted low-order polynomial along the LOS. And any special treatments to point sources have been sidestepped. In this method, the inverse-variance weighting scheme protects the frequency coherence of foregrounds, and most importantly, the polynomial function guarantees that only the spectrally smooth component of the point sources' radiation will be subtracted. One can therefore hope that most of the 21 cm signal emitted from the point sources will be preserved well in the polynomial-subtracted residuals. Furthermore, a key feature of this cleaning method is that the algorithm is blind, indicating that it is not necessary to resolve point sources in future observations. One can thus get a good start because the low-frequency BAO surveys will be performed with limited resolutions. Here, this formalism is employed to remove the radio foreground contamination. Refer to the original paper for more details of the cleaning method.

\section{Results}

In this work, we aim to study the feasibility of detecting BAOs in the 21 cm intensity fluctuations after reionization, using a radio interferometric array. The measurement uncertainties depend on many parameters, and we will mainly consider the impacts of telescope design, observation strategy and foreground subtraction techniques. To get a sense of the precision which we are able to achieve, we first describe a fiducial observation in Section 3.1, and further quantify its sensitivity in Section 3.2. A detailed account of the extent to which the various parameters matter is represented in Section 3.3.

\subsection{fiducial model}

Table 1 lists the parameter values adopted in our fiducial model. Most of these parameters reflect the current technology, and some specifications may be subject to change because of the developing experiments. Under the fiducial model, we perform the simulations described in \S2.2, and then use the proposed foreground cleaning method on the visibility data cube. Figure 4 illustrates the ability to remove the foreground contamination with instrumental response. The top row of the figure shows the simulated visibility cube. The left panel is a map from a single frequency channel, and the right panel plots a $\nu$-\emph{v} plane slicing along the frequency direction. Moreover, the bottom row shows the same cuts, but in the residual data cube following the polynomial subtraction. Qualitatively, the residual visibilities have been suppressed to a level of order $\sim 10$ $\mu${K}. This represents the fiducial scenario that can be achieved by our foreground cleaning method to the simulated interferometric measurements. In principle, the amplitude of the residuals could be made arbitrarily low by increasing the order of the polynomial used in the spectral fitting. However, the process of foreground subtraction may accidently remove the spectrally smooth component of the cosmological signal, and hence lead to suppression in the final estimate of power spectrum primarily on large scales. These artifacts have been discussed in previous work \citep[e.g.][]{Bowman09,Harker10,Mao11}. We should thereby perform the polynomial fit with order as low as possible.

%Fig.8
\begin{figure*}
\begin{center}
\includegraphics[angle=270, scale=0.7]{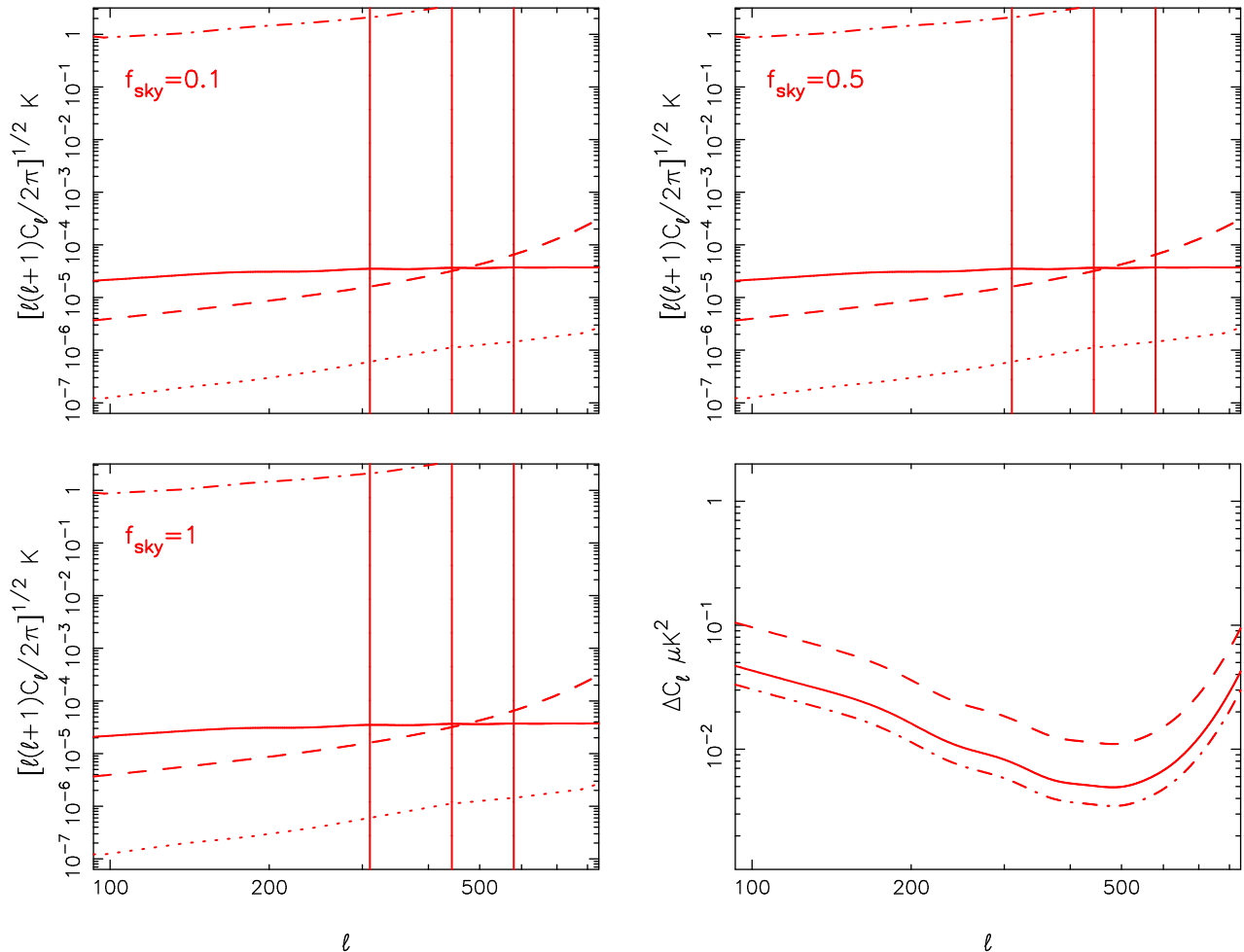}
\caption{Same as Figure 6, but for the effect of the covered sky area: sensitivity estimates for the fiducial model but with $f_{\rm sky}=0.1,0.5,1.0$. The bottom right panel shows the possible variances for $f_{\rm sky}=0.1$ (dashed line), $f_{\rm sky}=0.5$ (solid line), and $f_{\rm sky}=1$ (dash-dotted line).}
\end{center}
\end{figure*}

\subsection{sensitivity estimate}

After foreground subtraction, we are left with a residual data cube containing the fitting errors and the thermal noise. With this data set in hand, we can compute the analytic approximations of measurement uncertainties in a simple way that is often accurate enough for studying the experiment design. \citet{Knox95} has shown that the noise in a pixelized map could be modeled as a random field with an angular power spectrum given by
\begin{eqnarray}
C^{N}_{\ell}=\omega^{-1}e^{\theta^2_b \ell(\ell+1)}=\sigma^2_{\rm pix}\Omega_{\rm pix}e^{\theta^2_b \ell(\ell+1)},
\end{eqnarray}
where we take the pixel solid angle to be $\Omega_{\rm pix}=\theta_{\rm fwhm} \times \theta_{\rm fwhm}$, and $\theta_b$ is the standard deviation for a Gaussian beam function in the experiment. The most straightforward way of estimating the rms error in the map is to specify $\sigma_{\rm pix}$. For the observation we simulate here, $\sigma_{\rm pix}$ is defined as the standard deviation of the temperature fluctuations in the residual sky map, which has contributions from the residual foregrounds and other systematic effects. This approximation only strictly holds for a Gaussian distribution, and the foreground residuals in the sky map may not be a Gaussian field, but will be spatially correlated. However, we ignore these and use this relationship here since it allows considerable simplification. The estimate of the 21 cm power spectrum could therefore be described as a normal distribution with a variance \citep{Knox95}
\begin{eqnarray}
\Delta C_{\ell}=\sqrt{\frac{2}{(2\ell+1)f_{\rm sky}}}[C_{\ell}+C^N_{\ell}].
\end{eqnarray}
Thus the simple estimate of the accuracy with which the expected power spectrum could be measured directly depends on the sampling variance, cosmological signal and pixel noise. In our simulations, we never create the 21 cm signal. Instead we base our estimate of the 21 cm power spectrum on an analytical approach
\begin{eqnarray}
P^{3D}_{21}(z,k)=T^2_0 P_{\rm HI}(z,k)=T^2_0 b^2 P_{\delta\delta}(z,k),
\end{eqnarray}
in which $P_{\delta\delta}(z,k)$ and $b$ represent the matter power spectrum and bias factor respectively. The mean brightness temperature due to the 21 cm line can be written as \citep{Chang08,Seo10}
\begin{eqnarray}
T_0=0.3 \rm mK \left(\frac{\Omega_{\rm HI}}{10^{-3}}\right) \left[\frac{\Omega_m(1+z)^3+\Omega_{\Lambda}}{0.29(1+z)^3}\right]^{-1/2} \left[\frac{1+z}{2.5}\right]^{1/2}.
\end{eqnarray}
Also, the angular power spectrum $C_{\ell}$ is derived using the standard Limber's equation under the flat-sky approximation \citep{Kaiser92}.

%Fig.9
\begin{figure*}
\begin{center}
\includegraphics[angle=270, scale=0.7]{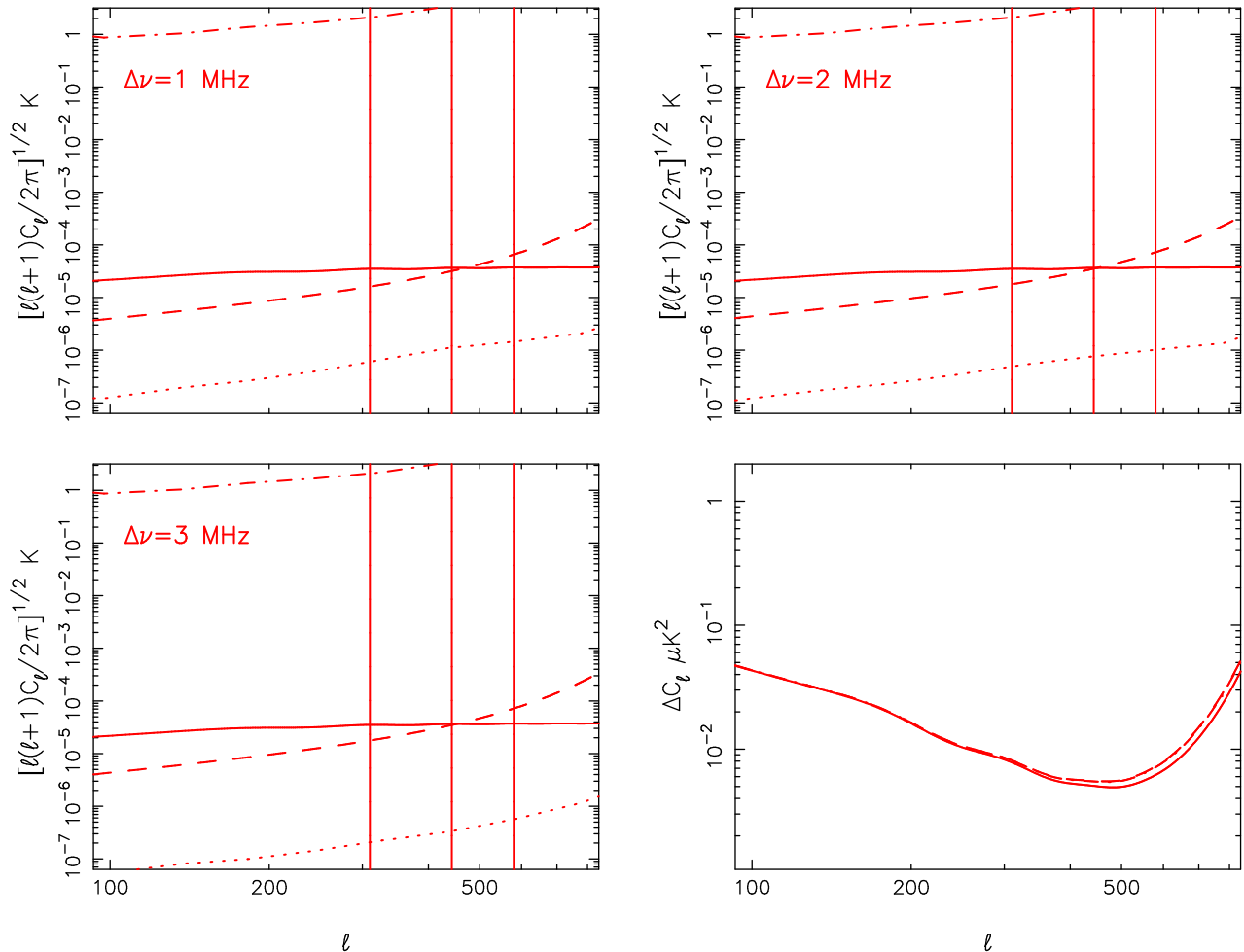}
\caption{Same as Figure 6, but for the effect of the frequency resolution: sensitivity estimates for the fiducial model but with the foreground subtraction performed by fitting three-order polynomials to the visibility spectra with various frequency resolutions. In the bottom right panel, the possible variances are plotted for $\Delta \nu=1$ MHz (solid line), $\Delta \nu=2$ MHz (dashed line), and $\Delta \nu=3$ MHz (dash-dotted line).}
\end{center}
\end{figure*}

In Figure 5, we show $C_{\ell}$, $C^N_{\ell}$ and $\Delta C_{\ell}$ for the fiducial model. The dashed line in the top panel represents the angular power spectrum of the intensity fluctuations in the foreground subtracted sky map, compared to the input foregrounds (dash-dotted line), detector noise (dotted line) and the expected 21 cm signal (solid line). The vertical lines indicate the locations of the first three BAO peaks. One can see that the huge contaminants will be effectively removed with obviously smaller residuals which are less than one part in $10^5$ of original model. Therefore the first two peaks of BAO wiggles could be measured with required precision. However, for $\ell>1/\theta_b$ the noise level increases rapidly. To a certain extent, this is induced by the fast growth of the thermal noise at small scales. In the bottom panel, the variance $\Delta C_{\ell}$ changes with $\ell$: it decreases at the first and then increases. Its minimum value appears at $\ell \sim 500$. From Equation 3, it is evident that the sampling variance as well as the cosmological signal and pixel noise will together determine the quantity of $\Delta C_{\ell}$. Their effects at different scales will be studied in the next section.

%Fig.10
\begin{figure*}
\begin{center}
\includegraphics[angle=270, scale=0.7]{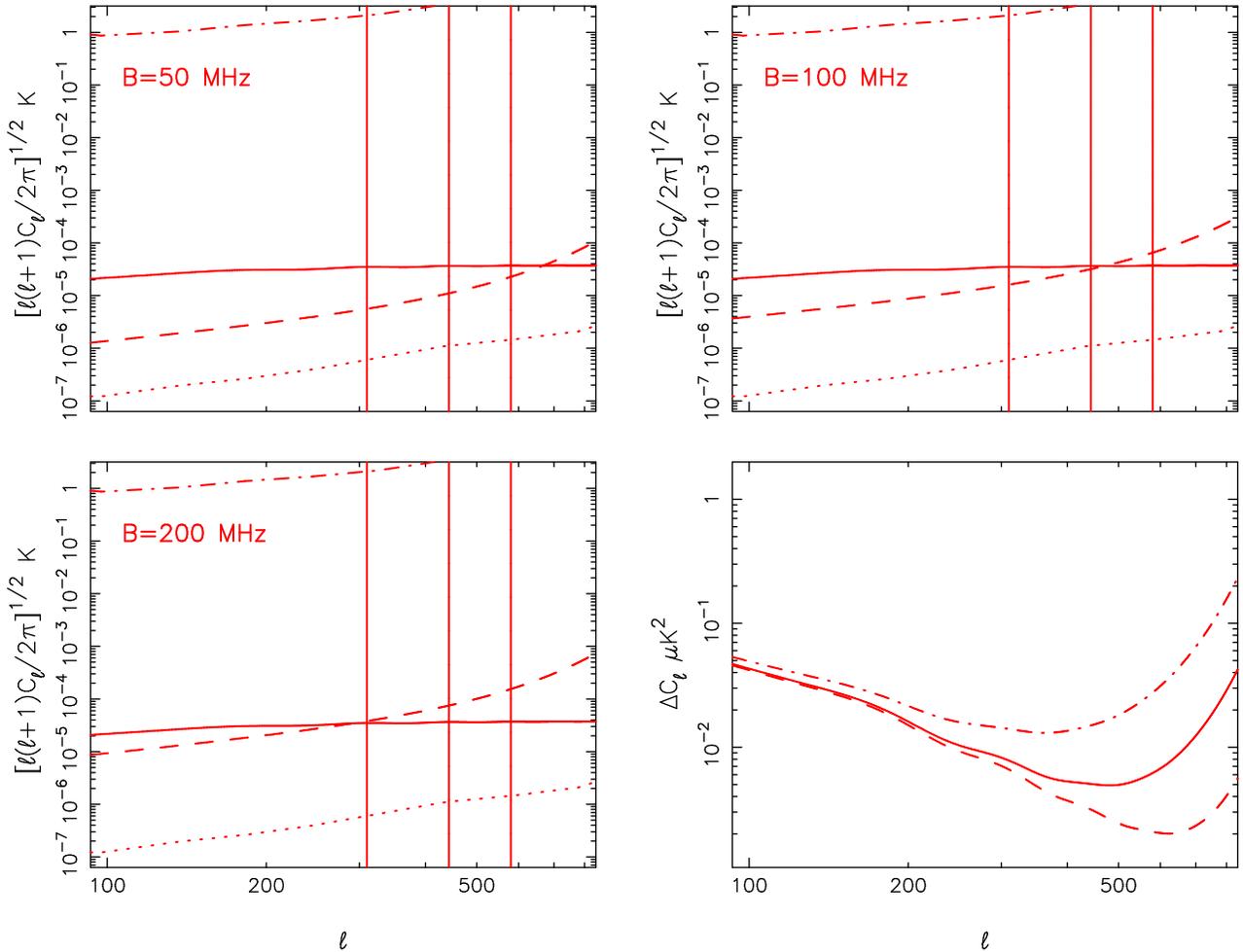}
\caption{Same as Figure 6, but for the effect of the bandwidth: sensitivity estimates for the fiducial model but with the foreground subtraction performed by fitting three-order polynomials to the visibility spectra over various frequency ranges. In the bottom right panel, the possible variances are plotted for $B=50$ MHz (dashed line), $B=100$ MHz (solid line), and $B=200$ MHz (dash-dotted line).}
\end{center}
\end{figure*}

\subsection{exploration of parameter space}

Above we investigated errors in the estimates of BAO scales under our fiducial model, including the effects of foreground contamination and instrumental response. Now we use a set of simulations with various parameters to investigate how well we could expect to measure the acoustic scales in future 21 cm surveys. In each case, we examine only one of these parameters and keep all others fixed.

\subsubsection{theoretical parameters}
Observations of quasar absorption spectra indicate that the intergalactic medium is highly ionized and a significant amount of neutral hydrogen could be found only in galaxies in the post-reionization universe. In this work, there is no need to describe the HI distribution, but rather we make a quantitative estimate of the total neutral hydrogen content and study its evolution at moderate redshifts. Our knowledge of the neutral hydrogen content is derived mainly from gas absorption in the HI hyperfine transition. These observations find that the comoving HI density in galaxies is almost constant with a density parameter of $\Omega_{\rm HI}=0.001$ beyond $z=1$, only doubling over the redshift range $z\approx2-4$ \citep{Peroux05,Wolfe05,Prochaska09}. Moreover, the proposed HI survey traces galaxies but not the underlying matter distribution, which over-weights the overdense regions and thus introduces an inherent bias. Since this bias factor is as yet fairly unconstrained observationally, we ignore its time and scale dependences and assume a constant value for it \citep[e.g.][]{Chang08,Wyithe09,Seo10}. This would be accurate enough for our purposes in the present work. Based on these, we use the halo model to calculate the mass power spectrum, and adopt $\Omega_{\rm HI}b=0.0001,0.0005,0.003$ for the pessimistic, fiducial and optimistic predictions in the theoretical model respectively. Figure 6 investigates these three scenarios. It appears to be true that the BAO measurements would benefit from increasing the intensity of 21 cm emission which is proportional to the cosmic density of neutral hydrogen in the universe. And in the bottom-right panel, the variances of power spectrum estimates increase with the amount of neutral hydrogen especially on large scales.

%Fig.11
\begin{figure*}
\begin{center}
\includegraphics[angle=270, scale=0.7]{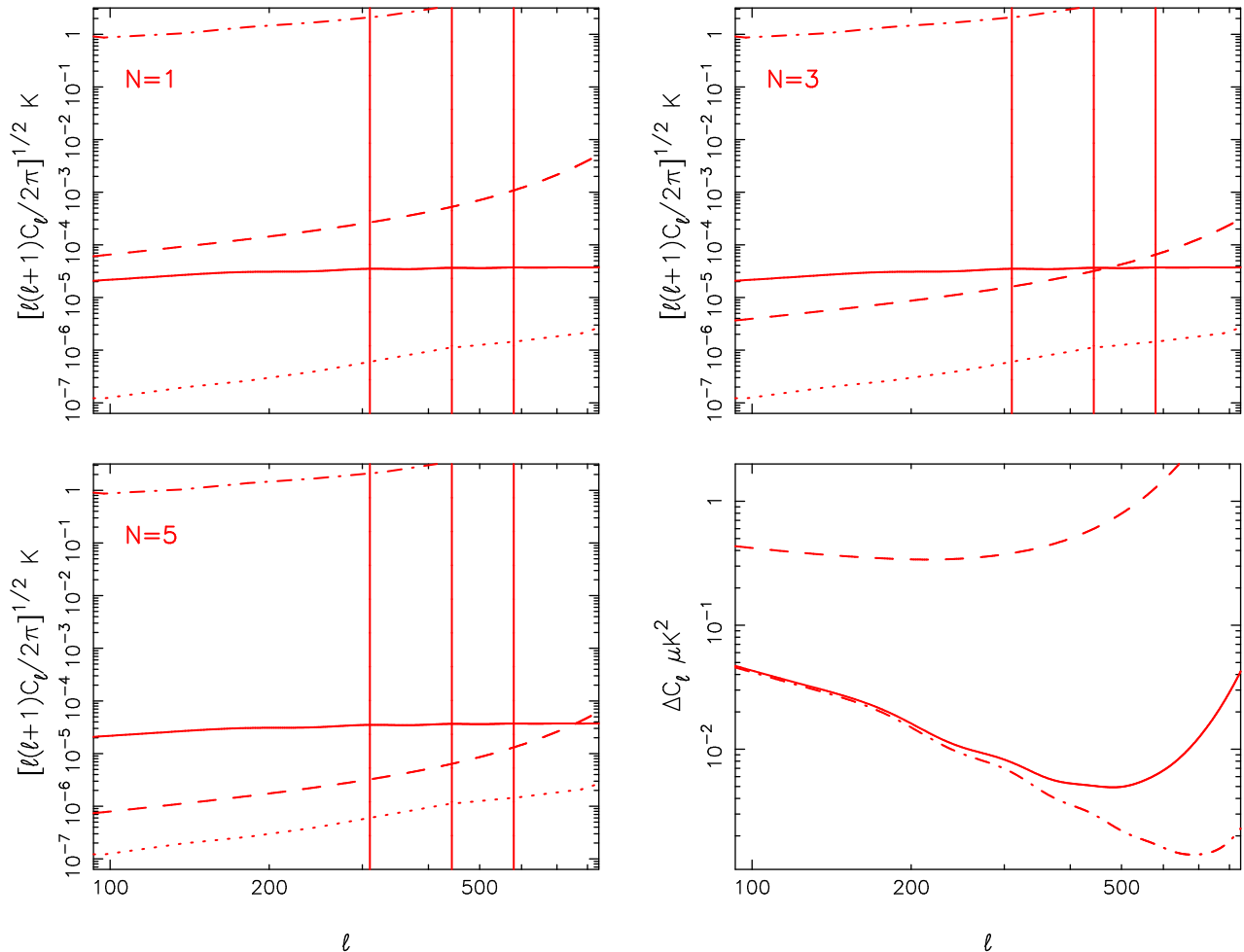}
\caption{Same as Figure 6, but for the effect of the order of the polynomial fit: sensitivity estimates for the fiducial model but with the foreground subtraction performed by fitting polynomials of various degrees to the visibility spectra. In the bottom right panel, the possible variances are plotted for $N=1$ (dashed line), $N=3$ (solid line), and $N=5$ (dash-dotted line).}
\end{center}
\end{figure*}

\subsubsection{instrumental parameters}
In a radio interferometric experiment, different distributions of dishes give different instrumental responses. We consider three cases here, in which dishes within diameters of 60, 100, 150 m are involved respectively and the total numbers of dishes are 226, 330, 487 correspondingly. Results are shown in Figure 7. In the top panels and the bottom-left panel, it is clear that the foreground subtraction can be done well using all these arrays, and there is no significant difference in performance at $\ell<300$. Nevertheless, one may note that the noise in the residual map (dashed lines) grows rapidly at large $\ell$ due to the exponential factor in the noise term. Meanwhile, the thermal noise (dotted lines) will increase faster for lower angular resolutions. In the bottom-right panel, for small values of $\ell$, $\Delta C_{\ell}$ is limited by the sample variance together with the cosmological signal, and thus decreasing the beam size $\theta_b$ decreases $\Delta C_{\ell}$ only slightly. But for $\ell \gtrsim 1/\theta_b$, the reduction in $\Delta C_{\ell}$ is dramatic, greatly expanding the available range of scales. For this reason, it is suggested that a BAO experiment should make measurements at resolution $\lesssim 10$ arcminutes in order to resolve the higher harmonics of BAO features in the extracted power spectrum. Moreover, the smaller beam experiment will provide information over a wider range of modes.

Now we focus on the effect of increasing the field of view. The observed fraction of the sky is defined by $f_{\rm sky}$. In Figure 8, we compare the possible errors in BAO measurements with $f_{\rm sky}=0.1,0.5,1$. Comparing the top panels to the bottom-left one, we do not find that the residual noise level (dashed lines) has reduced obviously. The most significant effect appears in the bottom-right panel, where the variance $\Delta C_{\ell}$ seems to be proportional to the factor $1/f_{\rm sky}$. The errors expected from the survey of a larger covering area appear smaller at most scales. It indicates that an experiment with poorer sky coverage will be less precise at every value of $\ell$, since the increase in the number of available modes decrease the sample variance errors. The figure also makes it clear that we are unable to extract meaningful information with our current methods until $f_{\rm sky}$ is large enough. For the BAO detections, it would be important to cover multiple sky areas in a fixed observing time, and hence reduce the large-scale errors which are dominated by sample variance.

\subsubsection{algorithmic parameters}
In this section, we examine three parameters relating to the foreground subtraction algorithm: frequency resolution, bandwidth of polynomial fit and order of the fitting polynomial. The results are plotted in Figure 9, Figure 10 and Figure 11 respectively.

From Figure 9, one can see that the lower frequency resolutions will make little difference in the measured power spectrum, indicating that the measurement uncertainties are not sensitive to the frequency resolution. This coincides with the conclusion presented in \citet{Mao11} which illustrated that the proposed foreground cleaning method would provide the residual data with similar quality even performed with various frequency resolutions.

In Figure 10, we investigate the effect of changing the frequency range over which the foreground spectra are fitted with smooth functions. At first sight, a smaller frequency bandwidth seems favorable for the BAO detections. After foreground cleaning, the overall level of noise (dashed lines in the top panels and the bottom-left panel) would increase with the bandwidth, and hence enlarge the variance $\Delta C_{\ell}$ especially at $\ell > 300$. However, we caution that the fewer frequency channels will degrade the signal-to-noise ratio in each band, furthermore, the cosmological signal on smaller scales may be accidentally removed during foreground subtraction. One should therefore approximate the frequency spectrum of foreground contaminants over a frequency range that is as broad as possible, with the constraint that the produced residuals lie below the expected level of cosmic signal. For real observations, a choice of the frequency range would be made in the context of a specific experiment.

Compared to the two parameters we mentioned above, the order of the polynomial fit will influence the results to the maximum extent, which is clearly shown in Figure 11. As we see in the top-left panel, there are insufficient degrees of freedom to remove radio foregrounds efficiently, and thus the rms noise (dashed line) in the residual maps may exceed the expected cosmological signal (solid line), inducing the destruction of measured BAO signatures. And obviously, the residual noise (dashed lines in the top-right and bottom-left panels) could be reduced significantly by increasing the order of polynomial used in foreground fits. Similarly, the bottom-right panel represents dramatic reductions in the detection variance. Nevertheless, as emphasized by previous work, some of the cosmological signal may be mistaken if N is too high \citep[e.g.][]{Furlanetto06,Liu09a}. In this work, we obtain believable measurements of the first two BAO peaks using a third-order polynomial to fit the real and imaginary parts of the visibility spectrum separately at each point in the \emph{uv} plane.

\section{Discussion}

Recently, it was realized that the redshifted 21 cm line emitted from neutral hydrogen can be used as a tracer of the large-scale structure even after reionization. Also, detection of BAOs on the inferred 21 cm power spectrum can be used to explore the geometry of the Universe especially in the dark energy dominated regime. In this paper, we quantified the likelihood of measuring BAO features near $z\sim 1$ with a set of ground-based radio interferometers. Simulations of the 21 cm interferometric measurements have been described, taking into account the detector noise, the foreground contamination and the frequency-dependent instrumental response. Different from galaxy redshift surveys in optical and radio wavebands, we developed a possible observing strategy without individual object detections. The foreground cleaning technology has also been studied, which is very different from both the CMB and EoR experiments. Moreover, we investigated what the accuracy of BAO measurements is likely to be available in future 21 cm surveys, and further discussed how the measurement uncertainties would change if different assumptions are made for key configurations of the radio telescope as well as some potential parameters used in the foreground subtraction method. The blind surveys of the redshifted 21 cm emission line have proven useful in offering information for acoustic scales at moderate redshifts, providing an important complement to the existing galaxy surveys.

In the fiducial scenario, we considered a pixel-by-pixel method to subtract continuum foregrounds from simulated visibilities, and found that a three-order polynomial can be used to well approximate the smooth spectrum. And in the cleaned sky maps, the detector noise together with the residual foregrounds will contribute to the pixel noise. Just as we saw in Figure 5, this noise term (dashed line) could be suppressed to a level of order $\sim 10$ $\mu$K at $\ell<500$ after a whole year integration. The imprints of the first two BAO peaks are therefore possible to be resolved on the 21 cm angular power spectrum. Although BAO experiments may not require fine resolutions, we found that an angular resolution of $\sim 10$ arcminutes is necessary in order to identify the higher harmonics of the BAO feature on the extracted power spectrum. This is because the residual noise would increase exponentially with the angular resolution, preventing significant detections on large values of $\ell$. We note that, in the fiducial scenario, all the antennas point toward the NCP and continuously observe a fixed sky patch. Therefore, the integration time can be approximated by the observing time. However, the total observing time should be longer if we observe a different part of the sky, depending not only on the location of the interferometer but also on the sky coordinates of the patch being observed. Making a simplifying assumption that a single field of view can be observed for 6-8 hours in each day \citep{Datta09,Datta10,Sarkar11}, the required integration time (1 year) would be possibly achieved over a period of 3-4 years. In addition, the precision of BAO measurements is proportional to the survey volume. Fortunately, next generation radio telescopes will benefit from the rapid rate of increase of field-of-view and collecting area. Then the survey speed can be greatly accelerated, which is likely to be orders of magnitude higher than that achievable by optical spectrographs. Generally speaking, each radio interferometer should produce an instantaneous FOV of $\sim 100$ $\rm deg^2$, and possess a collection area on the order of $\sim 1000$ $\rm m^2$. Moreover, several tens of separate pointings could be used to map out a significant fraction (\emph{e.g.} $\sim1/2$) of the sky within a limited observation time.

We investigated here the role of blind 21 cm surveys in BAO measurements. Our results suggest that radio interferometers have the potential to become very competitive facilities for performing cosmologically interesting tests. While it is a challenging experiment to achieve constraints on the order of $1\%$ in the BAO measurements, these calculations expand our understanding of the effects of instrument design, observing strategy and data analysis techniques, offering a new angle for the future generation of neutral hydrogen surveys.

\acknowledgments{We thank the anonymous referee for valuable comments which have greatly contributed to this work. We also thank Xiang-Ping Wu and Junhua Gu for helpful discussions. We are grateful to Jingying Wang for providing the foreground data used here, and to James Wicker for carefully reading the manuscript. Support for this work was provided by the National Science Foundation of China (Grant No. 11003019), and the Ministry of Science and Technology of China (Grant No. 2009CB824904).}

\clearpage

\end{document}